\documentclass[a4paper,10pt]{article}

\title{Physical applications of a new method of solving the quintic equation}
\date{}
\author{Victor B\^{a}rsan
\\IFIN-HH, Str.Atomistilor 407, 077125 Magurele-Bucuresti, Romania
\\e-mail: vbarsan@theory.nipne.ro}

\begin{document}

\maketitle

\begin{abstract}
Some physical applications of the Passare-Tsikh solution of a principal quintic equation are discussed. As an example, a quintic equation of state is solved in detail. This approach provides analytical approximations for several problems admitting until now only numerical solutions.
\end{abstract}

\section{Introduction}
The quintic equation is not of central importance in physics, but in some cases it can play a significant role. The first example is provided by Euler's (1767) and Lagrange's (1771) researches on the three body problem, in the context of celestial mechanics, where, in a particular situation, the distance between two pairs of bodies satisfies a quintic equation  \cite{knob} . In modern physics, the equation of state satisfied by the order parameter describing a 3D liquid mixture is also a quintic one \cite{miron}. Recently, a quintic equation of state, for pure substances and mixtures, has been proposed, as a refined variant of the well-known van der Waals equation \cite{koz}. The pressure gradient of a fluid in a magnetorheological damper satisfies a quintic equation \cite{MR}. Other applications occur in the physics of molecules, theory of elasticity, etc. (see the references of \cite{kingJMP}). An interesting example is provided by the phenomenological theory of phase transitins, when a physical system is described by a sixth order Ginzburg-Landau expansion of the free energy \cite{LL}: 

\begin{equation}
 F=-fu+\frac{a}{2}u^{2}+\frac{b}{4}u^{4}+\frac{c}{6}u^{6}
\end{equation}
where $u$ is the order parameter and $f$ - an external field. In the most popular cases, $u$ is the polarization or magnetization, and $f$ - the electric or magnetic field. The equilibrum condition $\frac{\partial{F}}{\partial{u}}=0$ gives the equation of state:
\begin{equation}
 f=au+bu^{3}+cu^{5}
\end{equation}
However, more appropriate for a physical analysis is the inverse relation,
\begin{equation}
u=u(f)
\end{equation}
giving the dependance of the order parameter on the external field. In order to pass from (2) to (3), we have to solve a quintic equation.

Although the roots of a quintic equation are known from the late 1850s onwards, their expressions are very complicated, and consequently of limited use for analytic calculations. The goal of this paper is to provide a simpler approach to this problem. It is essentially based on a recent result obtained by Passare and Tsikh \cite{passare}.

The outline of this paper is the following. In the second section, we present the most popular methods of solving the quintic equation. The third one is devoted to the presentation of the Passare-Tsikh formula for a root of the principal quintic. This formula allows us to express the root of such an equation as a series expansion, without making use of a new, and cumbersome, Bring (or Tschirnhaus-Bring) transformation. In the next section, using a modern presentation of some old results, a Tschirnhaus transformation is constructed in detail (in several other papers, it is just postulated, so the reader has a limited understanding of its grounds). The fifth section describes how a solution of a quintic equation of state - characterizing a system defined by a sixth order Landau expansion - can be obtained. In the last section, the relevance of this approach, for ferroelectrics and ferromagnets, is analyzed. The possibility of obtaining analytic approximations for a quintic equation of state, representing a rafined variant of the well knownvan der Waals equation, is also discussed.

\section{An outline of the most popular methods of solving the quintic equation}

The first step in solving a ``general quintic'' 
\begin{equation}
 x^{5} +a_{4} x^{4} +a_{3} x^{3}+ a_{2} x^{2}+a_{1} x+a_{0} =0
\end{equation}
is to reduce it to a simpler form, which does not contain quartic and cubic terms - the so-called ``principal quintic'' $(a_{4} = a_{3} = 0) $ . This reduction is obtained using a Tschirnhaus transformation; in the next section, we shall introduce such a transformation in a constructive way, so we shall not give it, now, a formal definition. The next step is to solve this simpler ``principal quintic''. There are, essentially, two approaches to this problem. 

An approach is based on the ``Kiepert algorithm'' (1878), reformulated in modern terms by King and Canfield (\cite{kingJMP}, \cite{king carte}). It consists in transforming the ``principal quintic'' into the so-called ``Jacobi sextic'', whose roots can be expressed in terms of some Jacobi functions (theta functions); the roots of the ``Jacobi sextic'' can be then used in order to calculate the roots of the ``principal quintic'', through an inverse Tschirnhaus transformation. The disadvantages of this method are that the transformations involved are quite complicated and even the Jacobi theta functions - quite unfamiliar.

Another approach is to use a Bring transformation (a more subtle - and more cumbersome - Tschirnhaus transformation) in order to transform the ``principal quintic'' into a ``trinomic quintic'' ($a_{4} = a_{3} = a_{2}=0$; the commas intend to suggest this specific choice of coeficients, giving the simplest trinomic quintic; other choices are $a_{3} = a_{2} = a_{1}=0$, etc.) . The roots of the trinomic equation: 

\begin{equation}
 x^{5} -x-t=0
\end{equation}
can be expressed as generalized hypergeometric functions, $_{4} F_{3}$. The ``simplest'' root is given by (\cite{weisstein quintic}, eq. (38)):

\begin{equation}
 x_{1}=t _{4} F_{3} (\frac{1}{5}, \frac{2}{5}, \frac{3}{5}, \frac{4}{5}; \frac{1}{2}, \frac{3}{4}, \frac{5}{4}; \frac{5^{5}}{4^{4}} t^{4})=
\end{equation}
\begin{eqnarray*}
=t+t^{5} +10\frac{t^{9}}{2!}  +15\cdot 14 \frac{t^{13}}{3!} +20\cdot 19 \cdot 18 \frac{t^{17}}{4!} +...
\end{eqnarray*}
where we have used Eisenstein's variant \cite{Eisen} while writting the coefficients of the series expansion.

The other 4 roots are obtained in terms of simple combinations of $_{4} F_{3}$ functions (\cite{weisstein quintic}, eqs. (39-42)). The disadvantage of this method is that the Bring transformation is quite complicated.

The root (6) was re-obtained recently by Glasser \cite{glasser} and Perelomov \cite{perelomov}, using simple methods. These authors gave also, both for equations of lower $(\leq4)$ or higher $(>4)$ order, formulae for the roots, expressed as hypergeometric functions. 

A remarkable step forward was made by Passare and Tsikh, who were able to express a root of a ``principal quintic'' as a series expansion in its coefficients, avoiding, in this way, a cumbersome Tschirnhaus transformation. The main point of our paper is to take advantage of the simplicity of the Passare-Tsikh formula \cite{passare}.

\section{The Passare-Tsikh solution}

Recently, Passare and Tsikh \cite{passare} obtained the expression of a root of the principal quintic

\begin{equation}
 Bx^{5} + Ax^{2} +x +1=0
\end{equation}
as a series expansion:
\begin{equation}
 x_5 =-\sum _{j,k\geq0} (-1)^{k} \frac{(2j+5k)!}{j!k!(j+4k+1)!} A^{j} B^{k}
\end{equation}
The domain of convergence of this series is given by the condition:
\begin{equation}
 5^5\vert B \vert ^{2} -4^4\vert B \vert +108 \vert A \vert ^{5} -27\vert A \vert ^{4} +1600 \vert A \vert \vert B \vert -2250 \vert A \vert ^{2} \vert B \vert <0
\end{equation}
It is easy to check that, for $A=0$, the root (8) becomes the ``simplest'' root of the trinomic quintic (5); indeed, 

\begin{equation}
 _{4} F_{3} (\frac{1}{5}, \frac{2}{5}, \frac{3}{5}, \frac{4}{5}; \frac{1}{2}, \frac{3}{4}, \frac{5}{4}; \frac{5^{5}}{4^{4}} t^{4})=
\end{equation}
\begin{eqnarray*}
=\sum _{k\geq0}  \frac{(5k)!}{k!(4k+1)!} t^{4k}
\end{eqnarray*}
Also, for $B=0$, (8) becomes
\begin{equation}
 x_+ =-\sum _{j\geq0} \frac{(2j)!}{j!(j+1)!} A^{j} =\frac{-1+(1-4A)^{\frac{1}{2}}}{2A}
\end{equation}
which is a root of the quadratic equation
\begin{equation}
Ax^2+x+1=0
\end{equation}
The convergence condition gives $1-4 \vert A \vert >0$, equivalent to $1-4A>0$ (the last inequality is obviously fulfilled for $A<0$), i.e. the condition for $x_+$ to be a real root.

Let us also mention that Passare and Tsikh gave a simple formula for a general trinomic equation:
\begin{equation}
a_0+a_m x^m + a_n x^n =0,   n>m>0
\end{equation}
which can be written, without any loss of generality, as:
\begin{equation}
1+x^m + a x^n =0
\end{equation}
in the form:
\begin{equation}
x_n (a)=\sum _{k\geq0} \epsilon^{1+nk}\frac{\Gamma(1+\frac{nk}{m})}{\Gamma(1+\frac{1+(n-m)k}{m})}\frac{a^k}{k!}
\end{equation}
where $\epsilon$ is the $m-th$ root of $-1$. For $m=1$ and $n=2$, (15) becomes (11), the root of (12).

Another particular case, $m=3$ and $n=5$, corresponds, in the classical formulation of the Landau theory of phase transitions, to the situation $T=T_c$. Indeed, at the critical temperature, the coefficient $a$ in (1) vanishes, $a(T=T_c)=0$, and the trinomial equation describes the critical curve, which, in the case of a magnetic system, is $M=M(H)\mid _{T=T_c}$. So, the formula (15) allows an exact description of such critical behaviours.

\section{The Tschirnhaus transformation}

In a seminal work, published in 1683, Tschirnhaus proposed a method for solving a polynomial equation $P_n (x)=0$, of degree $n$, which consists, mainly, in transforming it into a simpler one, $Q_n (x)=0$, where the polynomial $Q_n(x) $ has a simpler form than $P_n(x)$, in the sense that the terms of order $n-1,n-2$, ... are removed. However, in practical cases, Tschirnhaus's transformation is in general extremely complicated and difficult to use.

Recently, Adamchik and Jeffrey \cite{AJ} have given a very useful formulation of Tschirnhaus' work (and, also, of Bring's (1786) and Jerrard's (1852) work), putting these contributions in a modern, simple and accessible form. In the rest of this section, we shall follow closely their paper, in order to reduce a quintic equation having the form (2), to a principal quintic.

Let us note by $x_i$, $i=1,...5$, the roots of the equation
\begin{equation}
x^{5} +a_{3} x^{3}+a_{1} x+a_{0} =0
\end{equation}
which is similar to (2) and can be obtained from the general quintic (4) putting $a_{4} =a_{2}=0$. Our goal is to reduce the eq. (16) to a ``principal quintic'':
\begin{equation}
y^{5} + b_{2} y^{2}+b_{1} y+b_{0} =0
\end{equation}
having the roots $y_k$. Let us define:
\begin{equation}
S_n = S_n (x_k) = \sum _{k=1}^{5} x_{k}^{n}
\end{equation}
These sums can be calculated iteratively, using Newton's formula:
\begin{equation}
S_n = -n_{5-n} - \sum _{j=1}^{n-1} S_{n-j} a_{5-j}
\end{equation}
with $a_j =0$ for $j<0$. For our particular case, when $a_4 =a_2 =0$,
\begin{equation}
S_1 =0, S_2 =-2a_3, S_3 =0, S_4 =2 a_{3}^{2}-4a_{1}, S_5 =-5a_{0} 
\end{equation}
Also,
\begin{equation}
S_6 =-S_4 a_3 -S_2 a_1, S_7 =-S_5 a_3 - S_2 a_{0}, S_8 = -S_6 a_3 - S_4 a_1 -S_5 a_0 
\end{equation}
\begin{equation}
S_9 =-S_7 a_3 -S_5 a_1 -S_4 a_0, S_10 = -S_8 a_3 -S_6 a_1 -S_5 a_0
\end{equation}

Following Tschirnhaus, we shall suppose that the roots $y_k$ of (17) are related by the roots $x_k$ of (16) through the transformation (quadratic Tschirnhaus transformation):
\begin{equation}
y_k = x_{k}^{2} +\alpha x_k +\beta
\end{equation}
Due to the particular form of the principal quintic:
\begin{equation}
S_1 (y_k) =S_2 (y_k) =0
\end{equation}
where $S_n (y_k)$ are defined similarly to (18). Also, 
\begin{equation}
S_3 (y_k) = -3b_2, S_4 (y_k) =-4b_1, S_(y_k)=-5b_0
\end{equation}
It is easy to find that:
\begin{equation}
S_1 (y_k) = \sum _{k=1}^{5} (x_{k}^{2} +\alpha x_k +\beta)=S_2 +\alpha S_1 +5 \beta 
\end{equation}
and
\begin{equation}
S_2 (y_k) = S_4 +(\alpha ^2 +2\beta)S_2 +5\beta
\end{equation}
Using (20), we can express the coefficients $\alpha, \beta$ as simple algebric functions of $a_1$ and $a_3$:
\begin{equation}
\alpha=(\frac{3}{5}a_3 -\frac{2a_1}{a_3}),\beta=\frac{2}{3}a_3
\end{equation}
Now, with (25), we can find $b_2$, $b_1$, $b_0$ as functions of $a_3$, $a_1$, $a_0$ using (28) and the following three relations:
\begin{equation}
S_3 (y_k) = S_6 + 3\alpha S_5 + 3(\alpha^2 +\beta)(S_4 +\beta S_2) +5\beta^3
\end{equation}
\begin{equation}
S_4 (y_k) = S_8 +4\alpha S_7 +(3\alpha^2 +2\beta)S_6 +4\alpha (\alpha^2 +3\beta)S_5
\end{equation}
\begin{eqnarray*}
+(\alpha^4 +6\beta^2 +12\alpha^2 \beta)S_4 +2\beta^2 (3\alpha^2 +2\beta)S_2 +5\beta^4
\end{eqnarray*}
\begin{eqnarray*}
S_5 (y_k)=S_{10} +5\alpha S_9 +5 (\alpha^4 \alpha^2 +\beta)S_8 +10\alpha (\alpha^2 +2\beta)S_7 +10\beta^2 S_6 
\end{eqnarray*}
\begin{equation}
+\alpha (\alpha^4 +20\alpha^2 \beta +30\beta^2)S_5 +5\beta(\alpha^4 +6\alpha^2 +6\alpha^2 \beta +2\beta^2)S_4
\end{equation}
\begin{eqnarray*}
+5\beta ^3 (2\alpha^2 +\beta)S_2 +5\beta^5
\end{eqnarray*}
So, we have been able to reduce eq. (16) to a principal quintic, whose coefficients $b_2, b_1, b_0$ are determined as functions of $a_3, a_1, a_0$. If we know the roots $y_k$ of the ``principal quintic'', we can easely find the roots of (17), using the inverse Tschirnhaus transformation defined by (23).

With these preparations, we can exemplify our approach, solving explicitely a quintic equation with physical relevance - the equation of state for a Ginzburg-Landau system with sextic anharmonicities.

\section{The solution of the quintic equation of state}

We shall consider the eq. (2), or:
\begin{equation}
u^5 +a_3 u^3 +a_1 u +a_0 =0
\end{equation}
which is a particular form of (4), with:
\begin{equation}
a_4 =0, a_3 =\frac{b}{c}, a_2 =0, a_1 =\frac{a}{c}, a_0 =-\frac{f}{c}
\end{equation}
We shall obtain its roots, following four steps.

In the first step, we shall transform (32) into a ``principal quintic'':
\begin{equation}
z^{5} + b_{2} z^{2}+b_{1} z+b_{0} =0
\end{equation}
using a Tschirnhaus transformation, as described in the previous section.
In the second step, we shall rescale eq. (34),
\begin{equation}
\frac{z^5}{b_0} +\frac{b_2}{b_0}z^2 +\frac{b_1}{b_0}z +1=0
\end{equation}
through the transformation:
\begin{equation}
z=\frac{b_0}{b_1}w
\end{equation}
to obtain the form (7), with:
\begin{equation}
B=\frac{b_{0}^{4}}{b_{1}^{5}}, A=\frac{b_{0} b_{2}}{b_{1}^{2}}
\end{equation}
In the third step, we shall use the Passare-Tsikh solution (8), to obtain a (real) root of the ``principal quintic'' obtained from (35) via the transformation (36), having the coefficients (37), as functions of the physical parameters entering in the Landau expansion (1). 

Let us describe in some more detail how we can calculate effectively the roots of (32). The inverse of the Tschirnhaus transformation (23) is obtained solving the equation:
\begin{equation}
x^2 +\alpha x + \beta -y = 0
\end{equation}
(to simplify the notation, the indices of $x$ and $y$ have been dropped out) or, with $x=cz$,
\begin{equation}
A_0 z^2 + z + 1 =0, A_0 = - \frac{\beta_0 +y}{\alpha^2}, c=-\frac{\beta_0 +y}{\alpha}, \beta_0 = - \beta
\end{equation}
According to \cite{AJ}, any of its roots can be used to construct the inverse transformation; coosing $z_+$ as this root, it can be written, of course, in terms of radicals, or as a series:
\begin{equation}
 z_+ =-\sum _{j\geq0} \frac{(2j)!}{j!(j+1)!} A_{0}^{j} 
\end{equation}
Expressing iteratively $y^5$ as given by
\begin{eqnarray*}
y^5 =- \frac {A}{B}y^2 - \frac{1}{B} y -\frac{1}{B}
\end{eqnarray*}
we can write:
\begin{equation}
 (\beta_0 +y)^j = b_{j0} +yb_{j1} +y^2 b_{j2} +y^3 b_{j3} +y^4 b_{j4}
\end{equation}
where, for $j\leq 4$, the form of the coefficients $b_{jn}$ is trivial; for $j=5$, 
\begin{equation}
 b_{50}=\beta_0^5 - \frac{1}{B}; b_{51}=5\beta_0^4-\frac{1}{B}; b_{52}=10\beta_0^3 -\frac{A}{B};
\end{equation}
\begin{eqnarray*}
 b_{53}=10\beta_0^2; b_{54}=5\beta_0
\end{eqnarray*}
and for $n\geq 5$, they can be obtained from the relation:
\begin{equation}
 b_{n+k} = M^k b_{n}
\end{equation}
where $b_m$ is a column vector with elements $b_{m0}, ... b_{m4}$ and $M$ - a matrix with the following non-zero elements:
\begin{equation}
 M_{nn}=\beta_0, M_{15}=M_{25}=-\frac{1}{B};M_{35}=-\frac{A}{B}; M_{21}=M_{32}=M_{54}=M_{45}=1
\end{equation}
So, the solution $z_+$ takes the form:
\begin{equation}
 z_+ =B_0 +yB_1 +y^2 B_2+y^3 B_3 +y^4 B_4
\end{equation}
\begin{equation}
B_n=-\sum _{j\geq0} \frac{(2j)!}{j!(j+1)!} \frac{(-1)^j}{\alpha^{2j}}
\end{equation}
It would be convenient to express each of $y^n, 2\leq n \leq 4 $, in a form similar to (8), i.e. $y^n=\sum a_{jk}^{(n)}$, with $a_{jk}^{(n)}$ given by simple formulas. However, author's attempts of obtaining such expressions failed. Probably, the first step in such an endeavour is to identify the Passare-Tsikh series with an already studied function, like (for example) the two-variable hypergeometric function.

In the fourth step, we can obtain the other roots of the ``principal quintic'', constructing the quartic equation whose roots are the four roots of the ``principal quintic'', different of $x_5$, (8). Using Viete's relations, we easily get from (32):
\begin{equation}
x_1 +x_2 +x_3 +x_4 =-x_5
\end{equation}
\begin{equation}
x_1 x_2 + x_1 x_3 +x_1 x_4 +x_2 x_3 +x_2 x_4 +x_3 x_4 =x_{5}^{2}
\end{equation}
\begin{equation}
x_1 x_2 x_3 +x_1 x_2 x_4 +x_2 x_3 x_4 +x_1 x_3 x_4 =-a_0 -x_{5}^{3}
\end{equation}
\begin{equation}
x_1 x_2  x_3  x_4  =a_0 +x_{5}^{4}
\end{equation}

So, $x_1, x_2,  x_3,  x_4 $ are the roots of the quartic equation:
\begin{equation}
x^4 -x_5 x^3 +x_{5}^{2}+(a_0 +x_{5}^{3})x +a_0 +x_{5}^{4}
\end{equation}

\section{Final comments and conclusions}
It would be interesting to solve the equation of state (2) or (32), in the case of a specific ferroelectric. As already mentioned, the coefficient $a_{1}$ has a linear temperature dependence, $a_0$ is proportional to the electric field $E$, and $a_3$ is a parameter; so, the coefficients $A,B$ of the Passare-Tsikh solution are polinomials in $T$ and $E$; each term in the series (8) is a polinomial in $T$ and $E$. In this way, we can obtain an analytical approximation for the exact equation of state. However, according to a recent compilation of the coefficients of the Landau expansion of the free energy \cite{chen}, for uniaxial ferroelectrics, like $Sr_{0.8} Bi_{2.2} Ta_2 O_9$, the sixth order term is not yet experimentally measured. A similar situation occurs in magnetism \cite{magnet}. However, in the case of fluids, the quintic equation of state allows a more accurate description of the substance near the critical and the saturation point \cite{koz}; until now, only numerical calculations have been done. Our solution provides analytic approximations, useful for a better understanding of the physical phenomena. It can be also used in order to extend Sanati and Saxena's analysis of the Landau theory \cite{saxena}, including sextic anharmonicities.

\textbf{Acknowledgements} 
The author is grateful to Dr. Victor Kuncser for useful discussions. This research has been financed from CNCSIS Project Idei 2008 no.953.

\end{document}